\documentclass[a4paper]{article}
\usepackage{hyperref}
\hypersetup{hidelinks,
	colorlinks=true,
	allcolors=black,
	pdfstartview=Fit,
	urlcolor=magenta,
	breaklinks=true}
\usepackage{multirow}
\usepackage{makecell}
\usepackage{color}
\usepackage{graphicx}
\usepackage{lipsum}
\usepackage{INTERSPEECH2022}

\newcommand{\mysec}[1]{\vspace{-2mm}\section{#1}}
\newcommand{\mysubsec}[1]{\vspace{-2mm}\subsection{#1}}

\title{Radio2Speech: High Quality Speech Recovery from Radio Frequency Signals}
\name{Running Zhao$^{1,3}$, Jiangtao Yu$^{2,3}$, Tingle Li$^{2,3}$, Hang Zhao$^{2,3,*}$, Edith C.H. Ngai$^{1,*}$}

\address{
    $^1$The University of Hong Kong, Hong Kong SAR, China \\
    $^2$IIIS, Tsinghua University, Beijing, China \\
    $^3$Shanghai Qi Zhi Institute, Shanghai, China}
\email{\href{https://zhaorunning.github.io/Radio2Speech/}{https://zhaorunning.github.io/Radio2Speech/}}

\begin{document}

\maketitle

\newcommand\blfootnote[1]{%
\begingroup
\renewcommand\thefootnote{}\footnote{#1}%
\addtocounter{footnote}{-1}%
\endgroup
}

\begin{abstract}
Considering the microphone is easily affected by noise and soundproof materials, the radio frequency (RF) signal is a promising candidate to recover audio as it is immune to noise and can traverse many soundproof objects.
In this paper, we introduce Radio2Speech, a system that uses RF signals to recover high quality speech from the loudspeaker. Radio2Speech can recover speech comparable to the quality of the microphone, advancing from recovering only single tone music or incomprehensible speech in existing approaches. We use \emph{Radio UNet} to accurately recover speech in time-frequency domain from RF signals with limited frequency band. Also, we incorporate the neural vocoder to synthesize the speech waveform from the estimated time-frequency representation without using the contaminated phase.
Quantitative and qualitative evaluations show that in quiet, noisy and soundproof scenarios, Radio2Speech achieves state-of-the-art performance and is on par with the microphone that works in quiet scenarios.
\end{abstract}
\blfootnote{*Corresponding Author}

\mysec{Introduction}
The microphone “listens” to ambient audio like the human ear, giving rise to application in human-computer interaction and security (e.g., eavesdropping). However, microphones only perceive the audio and does not possess the sensitivity and discrimination of the human ear, so it is less capable of resisting noise and irrelevant speech \cite{denoise,irrelevantspeech,multiparty}. Further, both the human ear and microphone will be ``deaf" when there are soundproof materials between the audio source and receiver. The instinct deficiencies of microphones hinder their use in noisy and soundproof scenarios. Recent researches have proposed to use other modalities of microphones that allow them to sidestep these shortcomings, such as laser \cite{lasermicrophone}, Lidar \cite{ladarmicrophone}, and visual microphones \cite{visualmicrophone}. Yet, they are susceptible to lighting conditions or opaque substances. 


Radio frequency (RF) based sensing systems are possible candidates to deal with all these issues in a holistic way. This is based on the fact that RF signals are insensitive to surrounding acoustic noise and lighting and can traverse occlusions \cite{rfpose,rfpose3d}. 
Recent advances in RF-based systems have leveraged those properties to recognize or recover audio. Preliminary works \cite{wihear,RaSSpeR,Silentmicrowave} leveraged RF signals to recognize several words or phonemes according to the movement of vocal organ. Furthermore, in \cite{Eavesdropping}, a RF eavesdropping system was proposed to recover audio from the loudspeaker. UWHear \cite{UWHear} presented a system to recover and separate sounds from multiple loudspeakers based on RF signals. However, these systems can only recover the single tone music and speech with simple sentences (e.g., ``one two three"). It is conceivable that speech (please note the speech here and later refers to speech in the spoken corpus) contains complex frequency components and irregular harmonics compared with them, and thus speech recovery is a difficult task. Recently, WaveEar \cite{waveear} tried to recover the human reading voice via radar. Although it achieves good quantitative results, the recovered speech is incomprehensible to listeners. RadioMic \cite{radiomic} used RF signals to recover speech from the loudspeaker and medium, but it suffers from the same problem. The core function of microphones is to record the audio as close to the original as possible so that it is intelligible to listeners. This principle should also be followed by the RF microphone even though it has unique advantages in noisy and soundproof scenarios. Therefore, using RF signals to recover high quality speech has remained intractable. 

\begin{figure*}[th]
  \centering
	\includegraphics[width=2.7in]{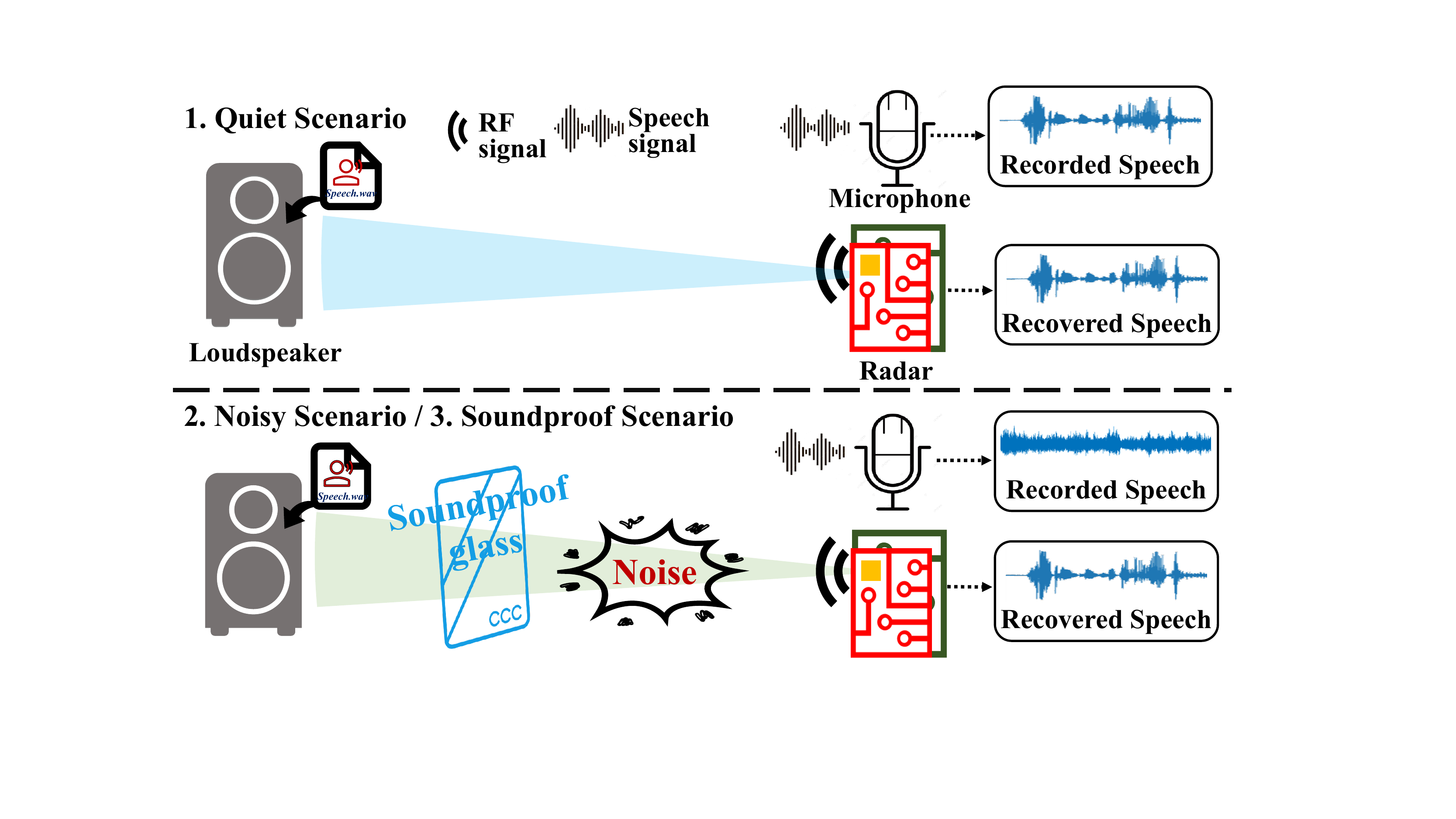}
	\hfil
	\includegraphics[width=3.65in]{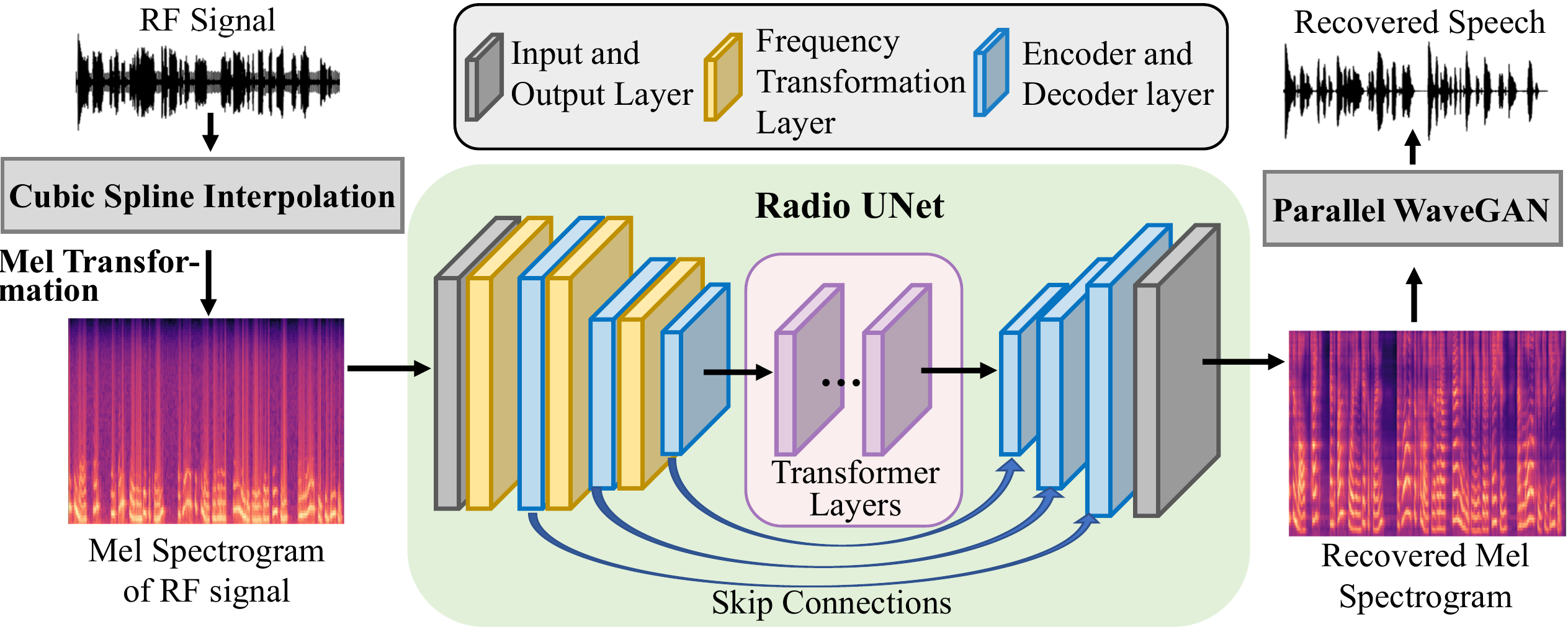}
	\hfil
  \caption{Illustration of Radio2Speech working scenarios (Left). In quiet scenario, Radio2Speech recovers the speech like the microphone. In noisy and soundproof scenarios, it performs the same as before, while the microphone fails. The architecture of Radio2Speech (Right). The input RF signal is upsampled to 8KHz and the Mel spectrogram of the RF signal is used as the input of Radio UNet.}
  \label{structure}
\end{figure*}

In this paper, we introduce Radio2Speech, a system that uses RF signals to recover speech of corpus with high quality from the loudspeaker. Nowadays, loudspeakers are ubiquitous in meeting room and home theater as a common audio source. Thus, it makes sense to acquire information from speech of loudspeakers.
As illustrated in Figure \ref{structure}, our system transmits the mmWave frequency-modulated continuous wave (FMCW) signal to the loudspeaker, and parses the reflected signal for recovery. In quiet scenario, Radio2Speech can recover high quality speech like a microphone. Even though in noisy and soundproof scenarios where microphone fails, our system is still able to recover high quality speech like in quiet scenario.

The distinctive properties of RF signals constrain the high quality speech recovery. First, restricted to device design, FMCW radar is not as sensitive as the diaphragm of microphones, and the upper frequency that RF signals can perceive is limited (below 1KHz). Also, its sampling rate is only 5.1KHz that is far lower than that of speech signals. Therefore, the high frequency band of RF signals is missing, which directly influences the speech quality. 
Second, RF signals contain much noise from FMCW radar hardware itself (referred to as phase noise \cite{phasenoise}) and environmental noise due to the multipath effect. Such noise contaminates the frequency and phase of the estimated speech signal. The above properties of RF signals have implications for speech quality, and proper handling of these properties is the key to recovering high quality speech.

Radio2Speech takes into account the above properties and deals with them as follows. Based on the UNet style network, we incorporate the frequency transformation layer to exploit correlation among all frequency harmonics, which have been proven to be helpful for time-frequency representation (TFR) reconstruction \cite{harmonic}. This incorporation allows the network to make full use of the limited frequency information to predict missing high frequency band due to limited frequency perception range and low sampling rate. Our UNet style backbone, named TUNet, employs Transformer layers at the bottleneck of convolutional neural network (CNN) based UNet due to its effectiveness in learning the contextual information in time-frequency domain.
The combination of them constitutes \emph{Radio UNet}.
Beyond this, we introduce a neural vocoder to synthesize speech signals from the recovered TFR, preventing the use of contaminated phase for synthesis. Experimental results show that in quiet, noisy and soundproof scenarios, Radio2Speech yields state-of-the-art quantitative and qualitative scores and is comparable to the microphone that works in quiet scenarios.

\mysec{Radio2Speech}

The general goal of Radio2Speech is to parse RF signals and recover high quality speech. Figure \ref{structure} describes the overall pipeline. First, the input RF signal is upsampled from 5.1KHz to 8KHz using the cubic spline interpolation that is normally used in bandwidth extension \cite{interpolation}. Then, Mel transformation is applied to the upsampled RF signal to obtain its Mel spectrogram. Next, it is fed into \emph{Radio UNet} to recover the Mel spectrogram of the speech signal. Finally, the neural vocoder, Parallel WaveGAN, is employed to reconstruct the natural speech waveform from the estimated Mel spectrogram.  

\mysubsec{Speech Recovery in Time-Frequency Domain}
We employ \emph{Radio UNet} to achieve recovery from the Mel spectrogram of RF signals to that of speech signals. Taking into account RF signal properties, \emph{Radio UNet} performs not only the transformation of Mel spectrograms, but also the speech bandwidth extension due to the low sampling rate and limited frequency perception range. 
Considering the Mel spectrogram pair of RF and speech signals $(M_{RF}, M_{s})$ fed to our network as input and target, the \emph{Radio UNet} acts as a function $T(\cdot)$ to perform the Mel spectrogram recovery $\widetilde{M}_{s} = T(M_{RF})$. 

In contrast to prior works using CNN based UNet as backbones in the TFR prediction task, our TUNet has the ability to capture long-range dependencies and global context, which CNN does not possess due to its localized receptive fields. The Mel spectrogram has a strong correlation along the whole time and frequency axes, and thus capturing contextual information is beneficial for our task. 
Inspired by the success of Transformer based UNet in image segmentation \cite{transunet}, our TUNet employs Transformer layers at the bottleneck so that network leverages contextual information in time-frequency domain. Specifically, the feature map $x\in \mathbb{R}^{H\times W\times C}$ with spatial resolution $H\times W$ and $C$ channels extracted from CNN encoder are directly reshaped as the sequence of flattened patches $x_p\in\mathbb{R}^{N\times(P^2\cdot C)}$ to perform tokenization, where $N=HW/P^2$ is the number of patches, and $P$ is set 1 here. Then, the patches are projected into D-dimensional patch embeddings by a linear projection, and a learnable positional embedding is added to them. The resulting sequence is the input of stacked Transformer layers, followed by CNN decoder.

As mentioned earlier, the high frequency band of RF signals is missing. It is crucial to make full use of the limited frequency band of RF signals for high frequency prediction. Especially, harmonic correlation along the whole frequency axis has been proven to be helpful for time-frequency representation reconstruction \cite{harmonic}. Although the backbone TUNet has the ability to capture contextual correlation for prediction, it pays less attention to such harmonics. Thus, we introduce the frequency transformation block from \cite{phasen}, but simplify it as the frequency transformation layer (FTL) to capture harmonics from RF signals with limited frequency band. FTL consists of three stacked CNN layers, a fully connected layer used as a transformation matrix, and a CNN layer used for concatenation. 
Assuming the input features extracted from the stacked CNN layers at time step $t$ as $f_{in}(t)\in \mathbb{R}^{F\times C}(t=1,\dots,T)$, where $T$, $F$ and $C$ are the time (width), frequency (height) and channel dimension, and transformation matrix as $W_{tr}\in \mathbb{R}^{F\times F}$, the output features $f_{out}$ after applying transformation matrix on input features can be represented as $f_{out}(t)=W_{tr}\cdot f_{in}(t)$. The transformed features $f_{out}$ stacked along the time axis contain the global frequency correlations, and then they are concatenated with the input features $f_{in}$ by a $1\times1$ convolution to capture both global and local frequency correlations. The FTL is added after input layer and each encoder layer. This allows the network to make full use of global frequency in limited frequency band of RF signals, constructing the Mel spectrogram with high frequency.

The L1 loss between Mel spectrogram of ground truth ${M}_{s}$ and estimated Mel spectrogram $\widetilde{M}_{s}$ is used for training. 

\mysubsec{Speech Waveform Synthesis}
We use the neural vocoder, Parallel WaveGAN \cite{parallelwavegan}, to synthesize speech waveform. The estimated Mel spectrogram of speech signal $\widetilde{M}_{s}$ is fed into the trained vocoder to convert the frequency acoustic features into natural sounding speech. 

The use of this vocoder considers the RF signal property that the noise contaminates its phase. Phase has a significant effect on the perceptual quality of generated speech waveform \cite{phaseimportance}. The intuition is to directly use inverse Short Time Fourier Transform (iSTFT) on the phase of input spectrogram of RF signals and estimated spectrogram to recover speech. However, as mentioned earlier, the phase of RF signals is contaminated by noise, and thus there is a large gap between the phase of RF spectrograms and target speech spectrograms. Directly using such a phase for speech synthesis affects speech quality. Also, there is little structure in the phase spectrogram, making it difficult to predict the phase spectrogram while estimating the magnitude spectrogram. Considering that \emph{Radio UNet} recovers accurate Mel spectrogram with affluent speech features, synthesizing speech waveform from Mel spectrogram can effectively avoid such shortcomings. Among the many neural vocoders, we adopt Parallel WaveGAN to transform the recovered Mel spectrogram into a natural speech waveform due to its fast and small-footprint characteristics.

\begin{figure}[!t]
  \centering
  \includegraphics[width=3.1in]{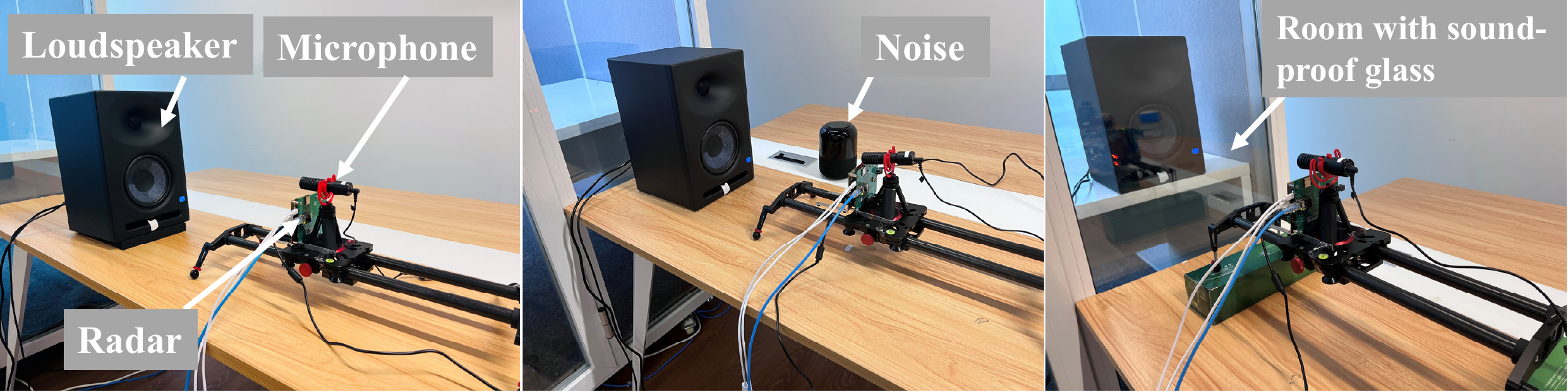}
  \caption{Experimental scenarios. From left to right are quiet, noisy and soundproof scenarios.}
  \label{scenarios}
\end{figure}

\mysec{Experiments and Results}
\subsection{Experimental Setup}
\noindent\textbf{Dataset.} We used two corpora, TIMIT \cite{TIMIT} for multi-speaker scenario and LJSpeech \cite{ljspeech} for single-speaker scenario, as the speech source for loudspeaker playback. LJSpeech consists of 13100 clips of a single speaker. The length of each clip varies from 1 to 10 seconds, and the total length is approximately 24 hours. TIMIT includes 6300 clips uttered by 630 speakers with different accents lasting 3.5 hours. The samples from such corpora are downsampled into 8KHz as the ground truth. 

We used a COTS mmWave FMCW radar TI AWR1642 and a data capture board DCA1000EVM to collect RF data corresponding to the above two corpora. After calibration, the length error between original speech and RF data is less than 10 ms. We conduct RF data collection experiments from three scenarios (see Figure \ref{scenarios}). (i) In the quiet scenario, the loudspeaker and radar were placed in a line and at a distance of 0.5m. (ii) The configuration of the noisy scenario was the same as the quiet scenario, but with an extra loudspeaker playing noise around the radar. (iii) While in the soundproof scenario, radar and loudspeaker were separated by soundproof glass, and the distance between them was 0.3m. In these scenarios, speeches were played from the loudspeaker at 85 dB SPL, similar to an actor's stage sound \cite{visualmicrophone}. Also, in these scenarios, a microphone was placed in the same location as the radar to collect data, which is used for comparison in Section \ref{results}. According to our investigation, the system performance decreases with increasing range (between radar and loudspeaker), similar to microphones. All because the signal (e.g., RF and audio signals) is attenuated as it propagates \cite{farfield}. The system performance also degrades with deviation in angle (with respect to radar) like unidirectional microphones, due to the limited perception angle \cite{unidirectional_microphone}. 

For the data from LJSpeech, the last 1000 samples were reserved for model evaluation and the remaining were used for training. Moreover, we randomly selected one sample from each speaker of the TIMIT, a total of 630 samples, for evaluation, and the others were used for training.

\noindent\textbf{Pre-processing.} Input RF signals were upsampled from 5.1KHz to 8KHz and target speech signals were downsampled to 8KHz before feeding into our network. 
Then, we extracted 80-dimensional log-Mel-spectrogram features with band-limited frequency range (60 to 4000 Hz). The window and hop sizes were set to 512 and 128.

\noindent\textbf{Implementation details.} TUNet has 3 encoder and decoder layers, 12 Transformer layers, and 1 input layer and output layer. All kernel sizes of CNN are set to $3\times3$. Also, CNN with stride size of $2\times2$ in encoder is used for downsampling, while decoder uses pixel shuffle to upsample feature maps. The configurations of FTL follow the settings in \cite{phasen}. The Transformer layers were pretrained on ImageNet \cite{imagenet}, and the Parallel WaveGAN were trained by the samples with 8KHz of the corresponding corpus. We use the normalized log-Mel-spectrogram with a fixed size of $80\times80\times1$ ($T\times F \times C$, about 1.28s) as the input for training efficiency. Our network was trained using the SGD optimizer with a learning rate of 0.01. 

\begin{figure}[!t]
	\centering
	\includegraphics[width=3.1in]{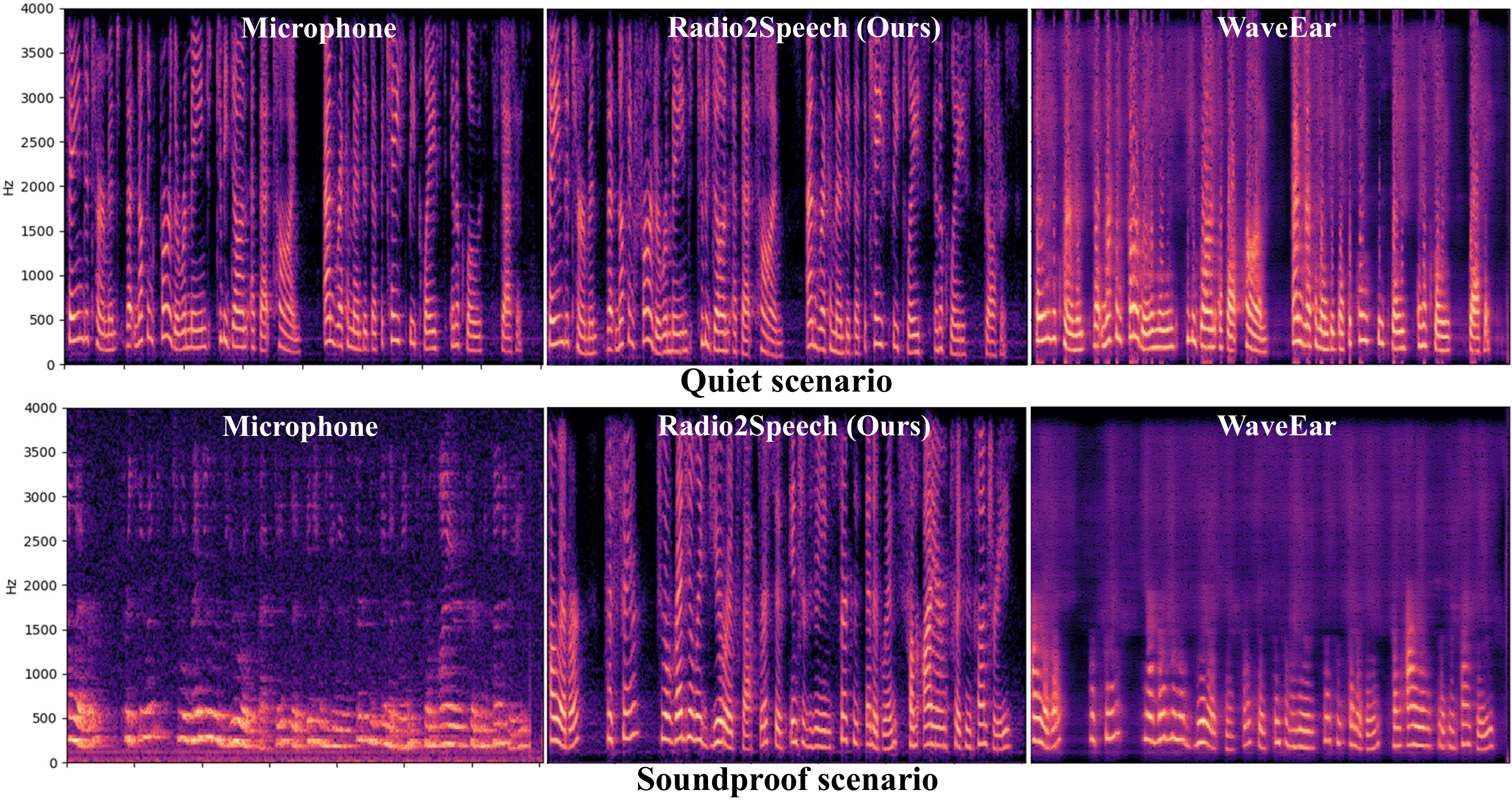}
	\caption{Spectrograms of recorded speech and recovered speech. The top and bottom rows correspond to quiet and soundproof scenarios, respectively.}
	\label{overall result}
\end{figure}

\mysubsec{Evaluation Metrics}
We use three quantitative metrics to evaluate the quality of recovered speech. \textbf{Short-time objective intelligibility (STOI)} measures the speech intelligibility (from 0 to 1). \textbf{Perceptual evaluation of speech quality (PESQ)} for narrow band is a perception evaluation related to subjective opinion (from 1 to 4.5). Signal-to-noise ratio (SNR) is frequently used to evaluate reconstructed speech quality. However, the SNR value of speech synthesized from the generative model is poor as its output is not exactly aligned with the target at sample level \cite{NUGAN}. In this case, SNR may not truly reflect the speech quality, thus, we choose \textbf{log-spectral distance (LSD)} that measures quality in frequency domain as the metric, and relevant parameters follow the settings in \cite{NUWave}. All metrics except LSD are better if higher.

We also use \textbf{mean opinion score (MOS)} to perform the subjective evaluation on Amazon Mechanical Turk. For each scenario, we randomly selected 10 RF data from the testing set of LJSpeech and TIMIT, respectively, a total of 20, and got the recovered speech from Radio2Speech and WaveEar. We also provided the corresponding speech recorded from the microphone. Each utterance was evaluated by 20 native speakers and the listeners were asked to rate a score from 1 to 5 (higher score means better quality) on speech naturalness.

\begin{table*}[!t]
  \caption{Quantitative evaluation results of ours, reference microphone (Mic) and WaveEar \cite{waveear} in three experimental scenarios with respect to LJSpeech and TIMIT. The best results are shown in bold. Since RF signals are not affected by the noise in our noisy scenario, the metrics are similar in noisy and quiet scenarios.}
  \label{objective results}
  \centering
  \footnotesize
  \begin{tabular}{c|c|ccc|ccc|ccc}
    \toprule
    \multirow{2}{*}{Dataset} &\multirow{2}{*}{Metrics} 
    
    &\multicolumn{3}{c|}{Quiet scenario} &\multicolumn{3}{c|}{Noisy scenario} &\multicolumn{3}{c}{Soundproof scenario} \\
    
    & &\textbf{Ours} & Mic & WaveEar \cite{waveear}  
    &\textbf{Ours}  &Mic & WaveEar \cite{waveear} 
    &\textbf{Ours}  &Mic & WaveEar \cite{waveear} \\
    \midrule
    \multirow{3}{*}{LJSpeech} 
    &LSD$\downarrow$  &\textbf{0.90} &1.05  &1.27  &\textbf{0.90}  &2.56  &1.27  &\textbf{1.02} &3.55 &1.44 \\
    &STOI$\uparrow$ &0.89 &\textbf{0.90}  &0.74  &\textbf{0.89}  &0.30  &0.74  &\textbf{0.75} &0.63 &0.62 \\
    &PESQ$\uparrow$ &2.50 &\textbf{3.00}  &1.78  &\textbf{2.50}   &1.24  &1.78  &\textbf{2.09} &1.59 &1.61 \\
    \cmidrule{1-11}
    \multirow{3}{*}{TIMIT} 
    &LSD$\downarrow$  &\textbf{0.97} &1.04  &1.23  &\textbf{0.97}  &2.78  &1.23  &\textbf{1.07} &2.24 &1.35 \\
    &STOI$\uparrow$  &0.80 &\textbf{0.92}  &0.68  &\textbf{0.80}  &0.29  &0.68  &\textbf{0.70} &0.59 &0.58 \\
    &PESQ$\uparrow$  &2.00 &\textbf{3.25}  &1.63  &\textbf{2.00}  &1.26  &1.63  &\textbf{1.71} &1.60 &1.43 \\
    \bottomrule
  \end{tabular}
\end{table*}

\begin{table}[!t]
  \caption{MOS results on speech naturalness. It is with 95\% confidence intervals.}
  \label{subjective results}
  \centering
  \footnotesize
  \begin{tabular}{c|ccc}
    \toprule
    Scenario &\textbf{Ours}   &Mic  &WaveEar~ \cite{waveear} \\
    \midrule
    Quiet       &3.10$\pm$0.16  &\textbf{3.21\boldsymbol{$\pm$}0.15} &2.77$\pm$0.18  \\ 
    Noisy       &\textbf{3.18\boldsymbol{$\pm$}0.15}  &1.21$\pm$0.17 &2.52$\pm$0.15 \\
    Soundproof  &\textbf{3.08\boldsymbol{$\pm$}0.14}  &1.43$\pm$0.15 &2.38$\pm$0.18   \\ 

    \bottomrule
  \end{tabular}
\end{table}

\begin{table}[!t]
  \caption{Ablation study on LJSpeech collected in quiet scenario.}
  \label{ablation}
  \centering
  \footnotesize
  \begin{tabular}{c|ccc}
    \toprule
    \multirow{1}{*}{Method} &LSD$\downarrow$  &STOI$\uparrow$  & PESQ$\uparrow$ \\ 
    \midrule
    \textbf{Radio2Speech (Ours)}  &\textbf{0.90} &\textbf{0.89} &\textbf{2.50}  \\
    w/o Transformer         &0.93 &0.86 &2.37   \\
    w/o freq trans layer    &0.98 &0.85 &2.21   \\ \cmidrule{1-4}
    w/o PWG+GriffinLim      &1.04 &0.82 &1.94    \\
    w/o PWG+iSTFT           &1.08 &0.77 &1.52    \\ 
    \bottomrule
  \end{tabular}
\end{table}

\mysubsec{Speech Recovery Results}\label{results}
We compare Radio2Speech with microphone and WaveEar \cite{waveear} in quantitative and qualitative evaluations. Some examples are available online.
Although WaveEar is designed to recover speech based on throat vibration, its network is also suitable for loudspeaker speech recovery and it outperforms other relevant methods. Thus, we re-implemented WaveEar consisting of encoder-decoder convolutional network and GriffinLim algorithm \cite{griffinlim} for comparison. For fair comparison, all methods was trained and evaluated by the identical training and test sets.

Figure \ref{overall result} uses spectrograms to visualize the recorded and recovered speech.  Since acoustic noise does not affect RF-based systems, the recovered spectrogram in noisy scenario is similar to that in quiet scenario, and we do not show it here due to space constraints. As can be seen, in quiet scenarios, Radio2Speech recovers comparable speech to the microphone, and it still performs well even in soundproof scenarios where the microphone completely fails. Moreover, Radio2Speech outperforms WaveEar in all scenarios.

\noindent\textbf{Quantitative Evaluation.} 
Table \ref{objective results} presents the quantitative results, evaluated by three metrics. 
It can be seen that in the quiet scenario, the metrics of Radio2Speech are almost as good as microphone for LJSpeech. Although ours performs slightly worse on the TIMIT dataset due to its limited samples and multi-speaker scenario, it is acceptable. Thus, Radio2Speech has a comparable ability with the microphone in terms of speech intelligibility and quality in the quiet scenario. Moreover, since RF signals are not affected by noise outside its narrow perception field, metrics of RF-based systems in noisy and quiet scenarios are the same. This indicates that Radio2Speech is still capable of recovering high quality speech in the noisy scenario. As for the soundproof scenario, although the existence of soundproof glass attenuates the RF signal power, metrics show that the speech recovered by Radio2Speech remains intelligible. As we expected, the microphone almost fails in noisy and soundproof scenarios, which is the instinctive advantage of RF microphones. These results demonstrate that Radio2Speech can recover speech comparable to that recorded by the microphone.

Radio2Speech outperforms WaveEar by a large margin across all three metrics for LJSpeech and TIMIT datasets in three experimental scenarios. Especially in the soundproof scenario, our metrics remain high, while those of WaveEar show its recovered speech is incomprehensible. Therefore, the intelligibility and quality of speech recovered from Radio2Speech are better than the state-of-the-art (SOTA) system, suggesting that our Radio2Speech can better exploit the proprieties of RF signals to recover high quality speech.

\noindent\textbf{Qualitative Evaluation.}
Table \ref{subjective results} shows the qualitative results of MOS evaluation, which can better show the actual perceptual quality of the recovered speech. As can be seen, in the quiet scenario, Radio2Speech achieves competitive performance compared to a microphone. In noisy and soundproof scenarios, our MOS scores remain high, which is on par with a microphone works in the quiet scenario. It concludes that Radio2Speech can recover high quality speech like a microphone in quiet, noisy and soundproof scenarios. Moreover, Radio2Speech surpasses WaveEar in all three scenarios, manifesting our system has better speech recovery ability than the SOTA system.
 
\mysubsec{Ablation Study}
Table \ref{ablation} reports the ablation study results, and all experiments were carried out on the testing set of LJSpeech collected in the quiet scenario.  
\emph{w/o Transformer} and \emph{w/o freq trans layer} represent that the Transformer layer and FTL are removed from Radio2Speech, respectively. All the metrics drop without Transformer, indicating the backbone of TUNet is better than UNet. By comparing \emph{w/o freq trans layer} to ours, we find that FTL provides 0.04 and 0.29 gain on STOI and PESQ, manifesting FTL can improve the speech quality by fully exploiting harmonic. Moreover, \emph{w/o PWG+GriffinLim} and \emph{w/o PWG+iSTFT} show the performance of using the GriffinLim algorithm and iSTFT to construct waveform instead of Parallel WaveGAN. Both of them result in 0.14-0.18 gain on LSD, 0.07-0.12 drop on STOI and 0.56-0.98 drop on PESQ. This demonstrates that the phase contaminated by noise significantly influences speech quality and the phase estimated by GriffinLim is not accurate. 


\mysec{Conclusions}
In this paper, we presented Radio2Speech, a system that parses RF signals to recover high quality speech from the loudspeaker. Experimental results show that in the quiet scenario, the performance of Radio2Speech is comparable to the microphone, and our system remains effective in noisy and soundproof scenarios where the microphone almost fails. Also, in these scenarios, Radio2Speech achieves SOTA performance in both quantitative and qualitative metrics.
We believe Radio2Speech could be utilized to perform more speech downstream tasks in the future.

\bibliographystyle{IEEEtran}

\bibliography{mybib}

\end{document}